\documentclass[twocolumn,showpacs,preprintnumbers,amsmath,amssymb]{revtex4}
\usepackage{tabularx,graphicx}

\usepackage{color}
\usepackage{hyperref}
\hypersetup{
    colorlinks=true,
    linkcolor=blue,
    filecolor=blue,      
    urlcolor=blue,
}

\begin{document}
%\documentstyle[aps]{revtex}
%\documentstyle[preprint,aps]{revtex}
%\begin{document}

\newcommand{\beq}{\begin{equation}}
\newcommand{\eeq}{\end{equation}}
\newcommand{\beqn}{\begin{eqnarray}}
\newcommand{\eeqn}{\end{eqnarray}}
\newcommand{\bmath}{\begin{subequations}}
\newcommand{\emath}{\end{subequations}}
\newcommand{\bra}[1]{\langle #1|}
\newcommand{\ket}[1]{|#1\rangle}

%\draft
\title{Corrigendum: ÔOn the dynamics  of the Meissner effectÕ 
(\href{http://iopscience.iop.org/article/10.1088/0031-8949/91/3/035801/meta}{2016 Phys. Scr. 91 035801},
 \href{http://arxiv.org/abs/1508.03307}{arXiv:1508.03307)}}
\author{J. E. Hirsch }
\address{Department of Physics, University of California, San Diego,
La Jolla, CA 92093-0319}

\begin{abstract} 
The paper erroneously assumed that the normal carriers giving rise to  the backflow could be either electrons or holes.
 \end{abstract}

\maketitle

  \setcounter{figure}{4}

     \begin{figure}
 \resizebox{8.5cm}{!}{\includegraphics[width=6cm]{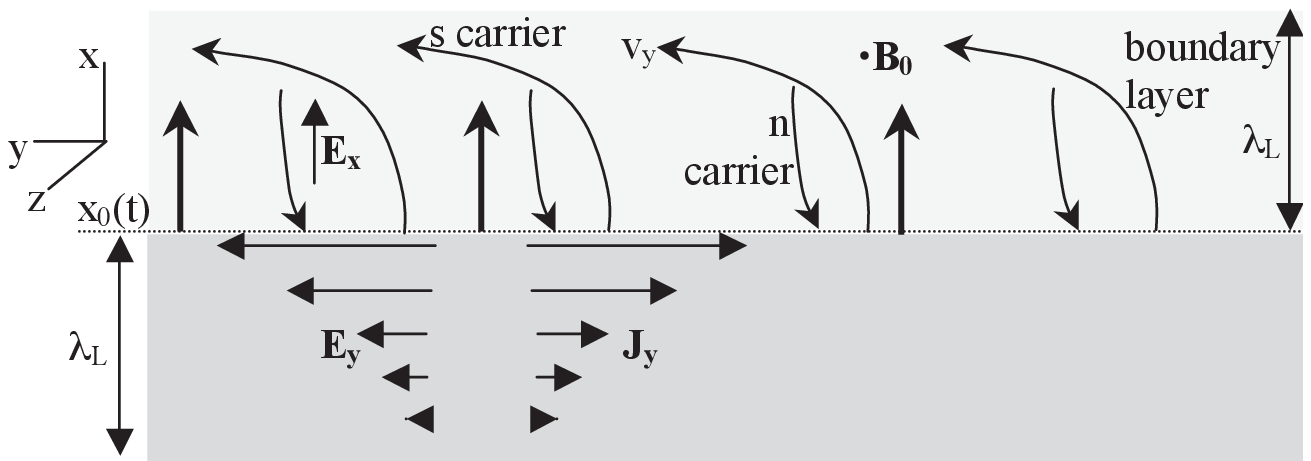}} 
 \caption { (original) As carriers become superconducting (s carriers) they thrust forward into the normal region over a boundary layer of thickness $\lambda_L$, and
 are deflected by the Lorentz force acquiring speed $v_y=-c/(4\pi n_s q \lambda_L)B_0$. This process creates an electric field $E_x$ in the $-\hat{x}$ direction that drives normal carrier (n carrier) backflow. Here, ``n'' stands both for ``normal'' and ``negative''. The normal carriers do not acquire a large $v_y$ in opposite direction because they scatter off
 impurities and transfer their $y-$momentum to the lattice.
 }
 \label{figure1}
 \end{figure}

   On page 7 the paper read:
 
  ``As the phase boundary advances at rate $v_0 = dx_0/dt$,
normal (negatively charged) carriers in the boundary layer are
backflowing at speed $v_0$ , or equivalently positive normal
carriers (holes) move forward together with the phase
boundary.
  Because normal carriers scatter off the lattice they
do not acquire a large $v_y$  from the action of the magnetic
Lorentz force; instead they transfer their y -momentum to the
lattice as a whole, thus accounting for momentum
conservation.''

The second half of that statement was incorrect. Similarly Figure 5 (reproduced below) and the last part of its caption that reads
``The normal carriers do not acquire a large $v_y$ in opposite direction because they scatter off
 impurities and transfer their $y-$momentum to the lattice.'' was not correct.

         \setcounter{figure}{4}
   
     \begin{figure}
 \resizebox{8.5cm}{!}{\includegraphics[width=6cm]{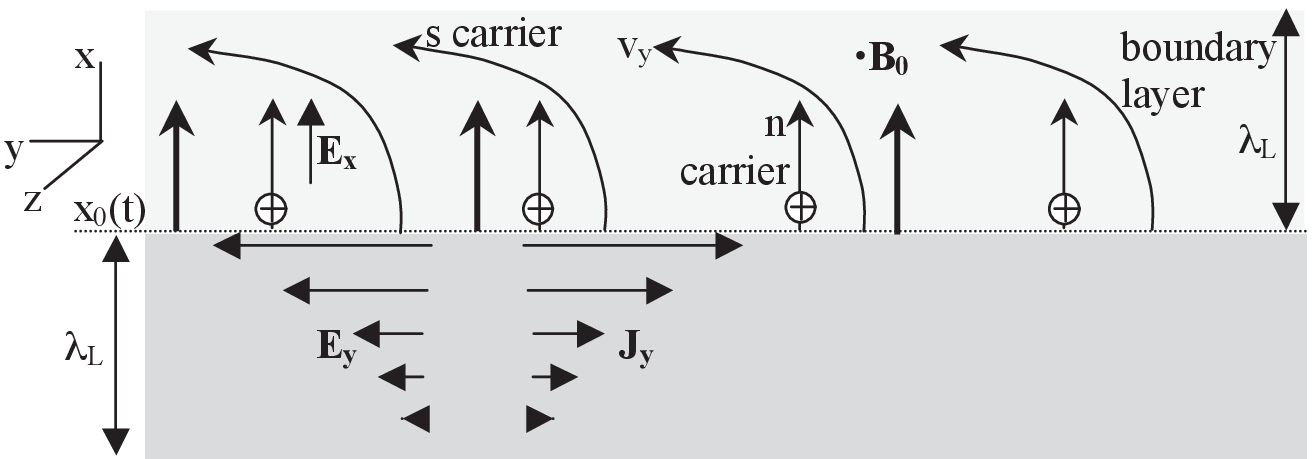}} 
 \caption { (corrected): As carriers become superconducting (s carriers) they thrust forward into the normal region over a boundary layer of thickness $\lambda_L$, and
 are deflected by the Lorentz force acquiring speed $v_y=-c/(4\pi n_s q \lambda_L)B_0$. This process creates an electric field $E_x$ in the $-\hat{x}$ direction that drives normal carrier (n carrier) backflow. Here, ``n'' stands for ``normal'' carrier. The normal carriers are $holes$ \cite{holesc} and they do not acquire any $v_y$ but instead
 propagate exactly along the $x$ direction, as explained in \cite{np1,np2,np3}, and in the process
   transfer  $y-$momentum to the lattice {\it without any scattering processes}.
 }
   \label{figure5}
 \end{figure}

   The paper was in error in ignoring the fact that within the theory of hole superconductivity the normal state carriers in a superconductor are necessarily holes \cite{holesc}. If the carriers are holes, electric and magnetic forces in the
   $y$ direction are exactly balanced and the backflowing carriers move exactly  in the $x$ direction, with no velocity component
   in the $y$ direction. The momentum transfer occurs through the flow of hole carriers and this accounts
   for momentum conservation {\it without scattering processes}, which is necessary to ensure the $reversibility$ of the process. For more details, see Ref. \cite{np1,np2,np3}.
   A corrected Figure 5 is given below.

      The author is grateful to an insightful referee who stated in his/her report:
      {\it ``for the situation at hand here, where the system is clearly a many-body one, the states being those of an interacting and hence correlated Fermion assembly, what exactly is meant by a `hole'? Again, and specifically in the context of promulgating a quite different view of a reasonably well accepted Ôstandard modelÕ it appears incumbent on the author to provide a cogent summary (a few sentences at the least) by way of elucidation of a `hole theory' actually is.''}
      
      The author responded to the referee's suggestion: ``In many papers listed in
 \href{http://sdphln.ucsd.edu/~jorge/hole.html}{physics.ucsd.edu/$\sim$jorge/hole.html}   the key issue of
holes versus electrons is addressed in detail, but it is not
relevant to this paper and there is no reason to address it
here.'' That answer was incorrect. The key issue of holes versus electrons was  highly relevant to this paper.

      The answer to the referee's query is contained in this erratum, which would have been  unnecessary if the author had
      heeded the referee's  suggestion to clarify ``what exactly is meant by a `hole' ''. The answer is, as illustrated in the
      corrected figure 5: {\it a hole is a normal state carrier  flowing in the direction of motion of the phase boundary in the reversible transition between normal
      and superconducting states. Because of its hole-like nature it propagates exactly perpendicular to the
      phase boundary without any velocity component in direction parallel to the phase boundary, which would
      render the transition irreversible in contradiction with experiment.}

\end{document}